\documentclass[aps,superscriptaddress,showpacs,nofootinbib,eqsecnum]{revtex4}

\usepackage{amsmath,epsfig,amsfonts,amssymb,graphicx,multirow}

\begin{document}

\title{Vector-like contributions from Optimized Perturbation in the
  Abelian Nambu--Jona-Lasinio model for cold and dense quark matter}

\author{Jean-Lo\"{\i}c Kneur}  \email{jlkneur@univ-montp2.fr}
\affiliation{CNRS, Laboratoire Charles Coulomb UMR 5221, F-34095,
  Montpellier, France} \affiliation{Universit\'e Montpellier 2,
  Laboratoire Charles Coulomb UMR 5221, F-34095, Montpellier, France}

\author{Marcus Benghi Pinto}  \email{marcus@fsc.ufsc.br}
\affiliation{Departamento de F\'{\i}sica, Universidade Federal de
  Santa Catarina, 88040-900 Florian\'{o}polis, Santa Catarina, Brazil}

\author{Rudnei O. Ramos} \email{rudnei@uerj.br}
\affiliation{Departamento de F\'{\i}sica Te\'orica, Universidade do
  Estado do Rio de Janeiro, 20550-013 Rio de Janeiro, RJ, Brazil}

\author{Ederson Staudt}  \email{tipller@yahoo.com} 
\affiliation{Departamento de F\'{\i}sica, Universidade Federal do Amap\'a, 
68902-280 Macap\'a, Amap\'a, Brazil}

\begin{abstract}

Two-loop corrections  for the standard  Abelian Nambu-Jona-Lasinio
model are obtained with the Optimized Perturbation Theory (OPT)
method. These contributions improve the usual mean-field and
Hartree-Fock results by generating a $1/N_c$ suppressed term, which
only contributes at finite chemical potential. We take the zero
temperature limit observing that, within the OPT, chiral symmetry is
restored at a higher chemical potential $\mu$, while the resulting
equation of state is stiffer than the one obtained when mean-field is
applied to the standard version of the model. In order to understand
the physical nature of these finite $N_c$ contributions, we perform a
numerical analysis to show that the OPT quantum corrections mimic
effective repulsive vector-vector interaction contributions.  We also
derive a simple  analytical approximation for the mass gap, accurate
at the percent level, matching the mean-field approximation extended
by an extra vector channel to OPT.  For $\mu \gtrsim \mu_c$ the
effective vector coupling matching OPT is numerically close (for the
Abelian model) to the Fierz-induced Hartree-Fock value $G/(2N_c)$,
where $G$ is the scalar coupling, and then increases with $\mu$ in a
well-determined manner.

\vspace{0.5cm}
\centerline{\large \it In press Int. J. Mod. Phys. E (2012)}  

\end{abstract}

\pacs{12.39.Fe,21.65.-f,11.15.Tk,11.15.Pg}

\maketitle

\section{Introduction}
\label{sec1}

Effective models are extensively used to understand the physics of
strong  interactions. This is particularly true in the study of the
structure of the  phase diagram of  Quantum Chromodynamics  (QCD),
specially in the low temperature (energy)  and high density region,
which is supposedly almost inaccessible through the present day
lattice techniques that make   direct use of QCD.  In this respect,
the use of effective models for  quark interactions, like for example
the Nambu-Jona-Lasinio (NJL) type of models \cite{njl}, has proven to
be extremely helpful to improve the  understanding of the phase
structure for strongly interacting matter.  

With NJL type of models the phase structure of QCD has been explored
mostly in  terms of the well known $1/N_c$ expansion, whose leading
contribution represents  the large-$N_c$ (LN) approximation and also
corresponds to  the mean-field-approximation (MFA) \cite{klevansky}.
Recently, the alternative Optimized Perturbation Theory (OPT) method
has been applied to the NJL model~\cite{njlsu2} and results  beyond
the MFA have been explicitly obtained for quantities related to the
QCD phase structure.  This application has been extended to the strong
coupling and  small current mass regime in order to investigate the
critical line on  the chemical potential-current mass
plane~\cite{strong}.

{} For example, it has been observed that, at $T=0$, the OPT predicts
that the first-order phase transition  takes place at a chemical
potential value that is higher than the one predicted by the MFA,
thus, chiral symmetry breaking is enhanced.  At this point it is
interesting to remark that this trend is also observed when the MFA is
applied to the NJL augmented by a repulsive vector-vector interaction
term,  $-G_V ({\bar \psi} \gamma^\nu \psi)^2$. It is then plausible to
imagine that the OPT $1/N_c$ correction captures some of the physics
associated with this kind of interaction,  which plays a major role at
finite densities. The aim of the present work is to address this
possibility by performing a numerical comparison between the OPT
results, obtained from the standard NJL model, and the MFA results,
obtained with the extended version of model, such that  the role of
the quantum corrections  generated by the former approximation can be
better understood. 

Although of academic interest only, the Abelian version of the NJL is
well suited for such a  task, since in this case, due to cancellation
of terms between scalar and pseudo-scalar  contributions at the
exchange  level \cite{klevansky},  only a $\mu$ dependent
contribution survives, allowing for a clearer comparison. Therefore,
the temperature does not play a very important role in the present
study and we, consequently, can restrict the analysis to the $T=0$
case.  By adjusting the value of the vector-vector coupling in the MFA
we will perform a numerical analysis exhibiting that the OPT results
can be approximately well reproduced, near criticality and quite
beyond, at  the value  $G_V \simeq G/(2 N_c)$, where $G$ is the scalar
coupling.   As we shall discuss, this could be anticipated by
comparing the OPT and MFA relations for the free energy density. We
then argue that one of the OPT main effects is to induce a new type of
physics within the standard version of the NJL, by radiatively
generating a vector-vector type of $1/N_c$ correction that mainly
affects the finite $\mu$ results.  The possible generalization of
these findings and their consequences for general four-fermion
effective models, such as the non-Abelian NJL model as well as  the
Gross-Neveu model, and perhaps even for  QCD, is currently being
investigated.  In this respect, the present investigation, that is
carried out in this simple case, is important in its own right, since
it offers the necessary framework to pursue such generalizations. We
should also notice that the particular value $G_V = G/(2 N_c)$ at the
level of our approximations just corresponds to the Fierz-induced
Hartree-Fock (HF) approximation, which is not surprising since the OPT
at first order  involves corrections topologically similar to Fierz
exchange terms. However, when considering the $\mu$-dependence, as we
shall illustrate the OPT  incorporates more corrections than simply
Fierz-exchange terms, and  the $G_V$ value best fitting near
criticality is merely a numerical  accident, only valid for the
Abelian model. An additional bonus of our  study is thus to produce a
simple general analytical comparison of the MFA  with vector-vector
interaction approximation with the OPT one, from which  we will be
able to show the differences and advantages of the OPT at  first-order
over the HF approximation. 

The remaining of this work is organized as follows. In Sec. \ref{sec2}  we
derive the free energy density using the OPT  formalism.  In Sec. \ref{sec3} we
perform a numerical comparison  between the OPT and the MFA results
for the thermodynamical quantities at $T=0$. The analytical comparison
of the OPT results with those coming from MFA with $G_V \ne 0$  is
studied in Sec. \ref{sec4}. Our conclusions are presented in Sec. \ref{sec5}.

\section{The OPT Free Energy for the Abelian NJL model }
\label{sec2}

The simplest version of the Abelian NJL model is described by a
Lagrangian density for fermionic fields given by~\cite{njl}

\begin{equation}
\mathcal{L}={\bar \psi}\left( i{\partial
  \hbox{$\!\!\!/$}}-m_{c}\right) \psi +G\left[ ({\bar
    \psi}\psi)^{2}+({\bar{\psi}} i\gamma _{5}\psi )^{2}\right] ,  
\label{Lnjl}
\end{equation}
\noindent
where $\psi$ in general represents a $N_{c}$-plet quark of just one
flavor with current mass $m_c$. Therefore, here, we consider the
one-flavor model with  (global) $U(1)$ symmetry. In the OPT
interpolation prescription \cite{linear} applied  in the case of
four-fermion theories \cite{prdgn2d,prdgn3d,njlsu2}, one starts by
deforming the original theory with the  replacements $m_c \to m_c +
(1-\delta)\eta$ and $G\to \delta G$ in  Eq. (\ref {Lnjl}) where
$\delta$ is the alternative perturbative expansion parameter that
determines  the order at which the  OPT expansion is carried out and
$\eta$ is an arbitrary mass parameter to be set in a variational way
through an appropriate optimization procedure at a given order in the
OPT. The typical optimization procedure  used in the many previous
applications and that we also adopt here, is the Principle of Minimal
Sensitivity (PMS), defined by \cite{pms}

\begin{equation}
\left. \frac{d\mathcal{P}^{(k)}}{d\eta }\right\vert_{\bar{\eta},\delta
  =1}=0\;,
\label{pms12}   
\end{equation}
where ${\cal P}^{(k)}$ is some physical quantity calculated up to
order $k$ in the OPT.  {}From the above prescriptions to build the
interpolated model and rewriting the quartic interaction in
(\ref{Lnjl}) through a Hubbard-Stratonovich transformation introducing
auxiliary fields, $\sigma$ and $\pi$, we obtain the interpolated
Lagrangian density in the OPT formalism,

\begin{equation}
\mathcal{L}=\bar{\psi}\left\{ i{\partial
  \hbox{$\!\!\!/$}}-(m_{c}+\eta) +\delta \left[ (\eta-\sigma)
  -i{\gamma }_{5}{\pi} \right]  \right\} {\psi } -\delta \frac{ 1 }{4
  G }\left( \sigma ^{2}+{\pi}^{2}\right)\;.
\label{delta12}
\end{equation}
Note from the interpolated theory Eq. (\ref{delta12}) that  the
original Lagrangian is recovered for $\delta= 1$, but at any finite
order $k$ in the OPT, any perturbative result evaluated with
Eq. (\ref{delta12}) becomes dependent on $\eta$. Fixing it through the
optimization procedure~(\ref{pms12}), new kinds of contributions
beyond MFA are embodied in the non-trivial optimal $\eta$-dependence.  

{}For the studies involving the thermodynamics and phase structure of
a given field theory model, the most appropriate quantity to be
optimized has been shown \cite{prdgn2d,prdgn3d,njlsu2} to be the
effective potential  (or free energy density), ${\cal F}$.  Applying
the optimization procedure (\ref{pms12}) to ${\cal F}$ evaluated to
some order $k$ in the OPT, we can then access the system's phase
structure in a nonperturbative way beyond MFA.  {}From the
interpolated Lagrangian density, Eq.~(\ref{delta12}),  we can
immediately read the corresponding {}Feynman rules and then perform a
perturbative expansion  of the free energy density in powers of
$\delta$.  Then, the free energy density in the $\sigma_c$ direction
(using that $\langle \sigma \rangle = \sigma_c$ and $\langle \pi
\rangle = 0$) reads \cite{njlsu2}

\begin{eqnarray}
{\cal F}^{\rm OPT} &=&\frac{\sigma_c^{2}}{ 4G } +  2i N_c \int
\frac{d^{4}p}{\left( 2\pi \right) ^{4}}{\ln }\left[ -p^{2}+
  (m_c+\eta)^2 \right] - 4 i \delta  N_c  \int \frac{ d^{4}p}{\left(
  2\pi \right) ^{4}}\frac{(m_c+\eta) \left( \eta -\sigma_c \right)
}{-p^{2}+(m_c+\eta)^2 } \nonumber \\ &-&8\, \delta G N_c  \left[ \int
  \frac{d^{4}p}{\left( 2\pi \right)
    ^{4}}\frac{p_{0}}{-p^{2}+(m_c+\eta)^2  }\right] ^{2} \;.
\label{delta5}
\end{eqnarray}
\noindent

All the  integrals in Eq. (\ref{delta5}) are to be interpreted in the
Matsubara's finite temperature formalism with

\begin{equation}
\int \frac{d^{4}p}{\left( 2\pi \right)^{4}} \equiv \frac{i}{\beta}
\sum_{n=-\infty}^{+\infty} \int \frac{d^3 p}{\left( 2\pi \right)^3
}\;,
\end{equation}
and quadri-momenta given as $p=(i\omega_n+\mu,{\bf p})$, where
$\omega_n=(2n+1)\pi T,\;n=0,\pm 1,\pm 2, \ldots$, are the Matsubara
frequencies for fermions.  By performing the sum over these
frequencies  (see e.g. Ref. \cite{njlsu2}), one obtains

\begin{equation}
{\cal F}^{\rm OPT}=\frac{\sigma_c^{2}}{4 G}-2N_c I_1 (\mu,T,\eta) +2
\delta N_c(\eta+m_c) \left( \eta-\sigma_c \right) I_2(\mu,T,\eta)  +
2\delta G N_{c} \:I_3^{2}(\mu,T,\eta) \;,
\label{landauenergy}
\end{equation}
\noindent
where  we have defined, for convenience, the following basic
integrals:

\begin{equation}
I_1(\mu,T,\eta) = \int \frac{d^{3}p}{\left( 2\pi \right) ^{3}}\left\{
E_p + T\ln \left[ 1+e^{-\left( E_{p}+\mu \right) /T}\right] +T\ln
\left[ 1+e^{-\left( E_{p}-\mu \right) /T}\right]  \right\} \;,
\label{defI1}
\end{equation}

\begin{equation}
I_2(\mu,T,\eta) = \int \frac{d^{3}p}{\left( 2\pi \right)^{3}}
\frac{1}{E_{p}}\left[ 1-\frac{1}{e^{\left( E_{p}+\mu \right)
      /T}+1}-\frac{1}{e^{\left( E_{p}-\mu \right) /T}+1}\right]\;,
\label{defI2}
\end{equation}
and

\begin{equation}
 I_3(\mu,T,\eta) =\int \frac{d^{3}p}{\left( 2\pi \right)^{3}} \left[
   \frac{1}{e^{\left( E_{p}-\mu \right) /T}+1}-\frac{1}{e^{\left(
       E_{p}+\mu \right) /T}+1}\right]\;,
\label{defI3}
\end{equation}
where $E_{p}^{2}={\bf p}^2+(\eta+m_c)^{2}$.   Here, as in
Ref. \cite{njlsu2}, only the divergent integrals occurring in  Eqs.
(\ref{defI1}) and (\ref{defI2}) are regularized by  a  sharp
non-covariant three-dimensional momentum cutoff, $\Lambda$, so that
the Stefan-Boltzmann limit is observed.  Note that the second term in
Eq. (\ref {delta5}) corresponds to a gas of free fermions, whose mass
has been dressed by $\eta$, while the third term represents tadpole
type of contributions, proportional to the quark condensate, $\langle
{\bar \psi} \psi \rangle$. The last term  in Eq. (\ref{delta5}), which
is proportional to  the square of the quark number density,  $\langle
\psi^+ \psi\rangle^2$, comes from the two-loop terms in a
next-to-leading order correction in the $1/N_c$ expansion, which can
be easily seen by redefining $G\to G/N_c$. This term is responsible
for generating the vector type of corrections.

In order to perform our evaluations, we need to consider the general
PMS equation (\ref{pms12}), which can be conveniently expressed  in
the form \cite{njlsu2}

\begin{equation}
 \left\{ \left(\eta - \sigma_c \right) \left[ 1 + (\eta+m_c) \frac
   {d}{d \eta} \right] I_2 + 2 G \, I_3 \frac{d}{d \eta} I_3
 \right\}_{\eta = {\bar \eta}}=0 \,.
\label{simplePMS0}
\end{equation}

One observes from Eqs. (\ref{landauenergy}) and (\ref{defI3}) that the
first $1/N_{c}$ correction\footnote{In the non-Abelian case there is
also  a $1/N_c$ contribution proportional to $I_2^2$, which vanishes
in the Abelian case \cite{njlsu2}.}  (represented by the last term)
is only relevant when $\mu \neq 0$, as it vanishes for  $\mu =0$.  In
the latter case, the PMS condition gives $\bar{\eta} =\sigma_c $,
merely reproducing the well known MFA results. Thus, the relevant
contribution arises when $\mu \neq 0$. Since we are mainly interested
in the phase structure of the model and on the effects of the OPT on
it, one basic quantity of interest is thus the thermodynamical
potential, $\Omega$, whose relation to the free energy is given by
$\Omega= {\cal F}({\bar \sigma}_c)$. The optimized order parameter,
${\bar \sigma}_c$, is determined from the gap equation generated by
minimizing ${\cal F}$ with respect to the classical field,
${\sigma}_c$.  {}From Eq.~(\ref{landauenergy}) we obtain that

\begin{equation}
 {\bar \sigma}_c = 4 G  N_c ({\bar \eta}+m_c) I_2(\mu,T,{\bar
   \eta})\;,
\label{simpleGAP}
\end{equation}
\noindent
which can be inserted in the PMS equation to determine the optimized
pressure.

\section{Matching the OPT corrections in terms of a vector interaction}
\label{sec3}

The effects due to the OPT first-order corrections for cold and dense
quark matter can now be studied for the case of finite density effects
($\mu \neq 0$) at zero temperature which is easily achieved by taking
the limit $T \to 0$ in the  thermal integrals\cite{njlsu2}.
Physically,  this situation is relevant in studies related to neutron
stars for example. This is also the regime where lattice techniques
face more problems  with the sign problem.  Having obtained the
optimized pressure one can easily obtain the quark density number,
$\rho=dP/d\mu$, and  by taking into account both the gap and PMS
equations. The energy  density is given as usual by
$\varepsilon=-P+\mu \rho$ and allows us  to obtain the equation of
state (EoS).

Now, in order to carry out comparisons with the MFA let us recall
that, as emphasized in Refs.~\cite{koch,buballa_stab,buballa}, the
introduction of a repulsive vector-vector interaction term of the form
$-G_V( {\bar \psi} \gamma^\nu \psi)^2$ in Eq. (\ref{Lnjl}) is also
allowed by  chiral symmetry. Such a term can become important at
finite densities, generating a saturation mechanism, depending on the
vector coupling strength, that provides better matter stability. Also,
as far as phase transitions are concerned, it can influence the size
of the first-order  transition region and, hence, the location of the
(tri)critical point. As already mentioned,  the main goal of the
present work is to compare the OPT results obtained at $G_V=0$, with
the MFA results for $G_V \neq 0$ and whose free energy density is
given by \cite{buballa}

\begin{equation}
{\cal F}^{\rm MFA}=\frac{(M_{\rm MFA}-m_c)^{2}}{4 G}-2N_c I_1 ({\tilde
  \mu},T,M_{\rm MFA}) -4 G_V N_c^2 \:I_3^{2}({\tilde \mu},T,M_{\rm
  MFA})\;,
\label{landauenergyMFA}
\end{equation}
where $M_{\rm MFA}$ is the mass gap in MFA. In this case, one has to
simultaneously solve the equations

\begin{equation}
 \tilde{\mu}=\mu -4 G_V  N_c I_3({\tilde \mu},T,M_{\rm MFA}) \,\, ,
\label{mushift}
\end{equation}
and 

\begin{equation}
M_{\rm MFA}=m_c + 4 G N_c M_{\rm MFA} I_2({\tilde \mu},T,M_{\rm MFA})
\,\,.
\label{Mmc}
\end{equation}

{}For our purpose, it is crucial to notice that the {\it combined}
effects of $\tilde \mu$, defined by Eq. (\ref{mushift}),  and of the
$-4 G_V N_c^2 \:I_3^{2}$ term in  Eq. (\ref{landauenergyMFA}), is to
produce a net {\it positive} contribution  approximately $\sim 4 G_V
N_c^2 \:I_3^{2}$  (for $\mu \gg M$) to the the free energy
\cite{Fukushima1,Fukushima2}.  This last term turns out to be of the
same form as the OPT finite $N_c$  correction term,  $2G N_c\:
I_3^{2}$, so that one may guess that results obtained with this
approximation should be close to those obtained with the MFA at $G_V$
values of the order $G/(2 N_c)$, but not the same (see detailed
discussion in Sec. \ref{sec4}). This is precisely what we will check now in a
numerical fashion by considering the EoS. \\ {}Following the standard
procedure for the NJL model,  we set the OPT parameter values to
$\Lambda= 620 \, {\rm MeV}$, $G \Lambda^2=4.44$ with $m_c = 5.0\, {\rm
  MeV}$, so that at $T=0$ one matches by convention the ``physical''
values of $f_\pi=92.4 \, {\rm MeV}$, $m_\pi = 135 \, {\rm MeV}$, while
predicting $M =341 \, {\rm MeV}$,  and $- \langle {\bar \psi} \psi
\rangle^{1/3}= 191 \, {\rm MeV}$ \cite{ederson}. One subtlety is that
the $m_c$ value thus determined for OPT is strictly speaking slightly
different from the corresponding MFA $m_c$, when obtained consistently
from fitting the same pion mass value: this is due to the fact that
the OPT gives $G^2$ correction to the relation between $m_c$ and
$m_\pi$ \cite{njlsu2}.  More precisely, one finds  $m^{MFA}_c\sim
5.36\, {\rm MeV}$ instead of $m^{OPT}_c\sim 5.0\, {\rm MeV}$ for the
same $\Lambda=620 \, {\rm MeV}$ input. (In contrast the $G$ value
being obtained from $f_\pi$ is not changed from MFA to OPT,  since for
the $U(1)$ model,  $f_\pi$ receives no OPT corrections at $G^2$
order).  Within the MFA, we will also use the following representative
values for the vector coupling, $G_V = 0, \; 3G/(8 N_c), \; G/(2 N_c)$
and $3G/(4 N_c)$.

\begin{figure}[tbh]
\vspace{0.5cm} \centerline{\psfig{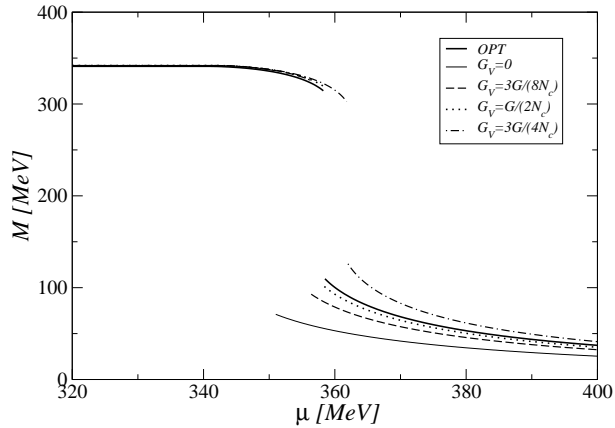}}
\caption{Effective quark mass (at $T=0$) as a function of the chemical
  potential.}
\label{massa}
\end{figure}

{}Figure \ref{massa} shows the quark effective mass  as function of
the chemical potential $\mu$, obtained with the OPT and the MFA for
the four different $G_V$ values. {}First, one notes that the $\mu_c$
value at which chiral symmetry is restored, via a first order
transition, increases with $G_V$ in the MFA. The critical chemical
potential value predicted by the MFA with $G_V= G/(2N_c)$  agrees the
better with the OPT prediction, as we can see also from the results
shown in Tab. \ref{tab}.

\begin{table}[htb]
\begin{center}
\caption{\label{tab} The critical  chemical potential $\mu_c(T=0)$ in
  the OPT and in the MFA (for the different values of $G_V$).}
\begin{tabular}{c|c|c|c|c|c}
\hline $G_V$ &  $0$ & $3G/(8N_c) $ & $ G/(2N_c) $ &$ 3G/(4N_c)  $ &
OPT ($G_V=0$)     \\ \hline $\mu_c \,({\rm MeV})$& 350.9& 356.3 &
358.3 & 361.9 & 358.5    \\ \hline
\end{tabular}
\end{center}
\end{table}

\begin{figure}[tbh]
\vspace{0.5cm} \centerline{\psfig{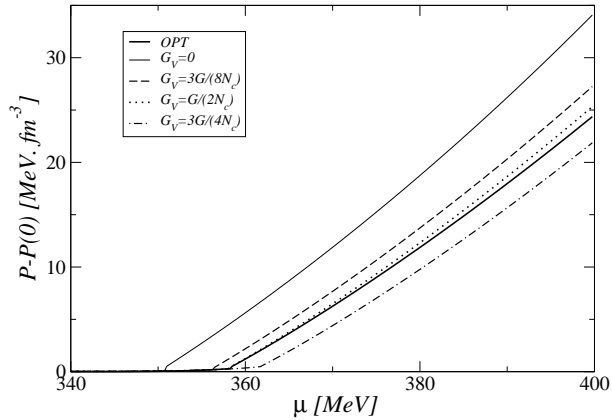}}
\caption{The pressure (subtracted by the value at $\mu=0$)  as a
  function of the chemical potential.}
\label{pressao}
\end{figure}

\begin{figure}[tbh]
\vspace{0.5cm} \centerline{\psfig{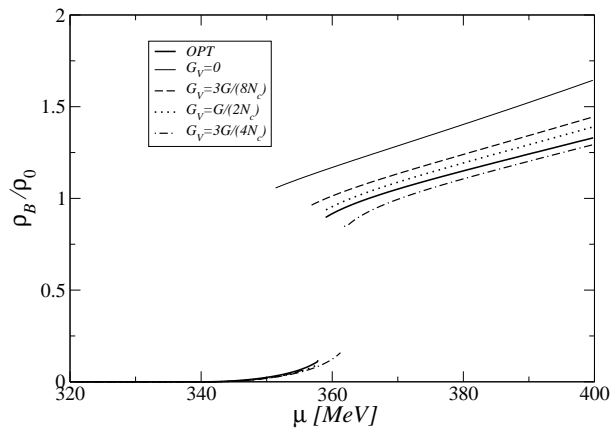}}
\caption{Baryon number density, $\rho_B=\rho/3$, in units of nuclear
  matter density ($\rho_0 = 0.17 \, {\rm fm}^{-3}$), as a function of
  the chemical potential.}
\label{densidade}
\end{figure}

\begin{figure}[tbh]
\vspace{0.5cm} \centerline{\psfig{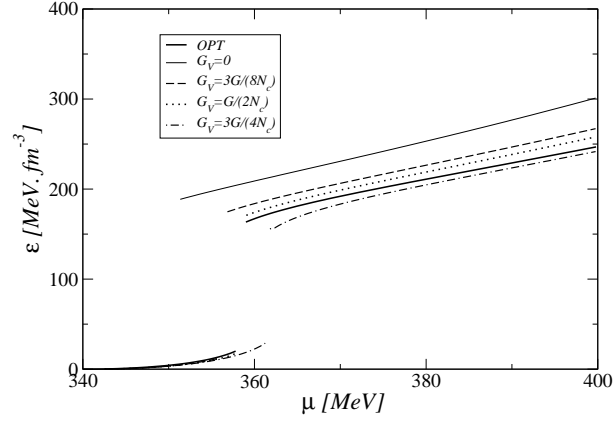}}
\caption{Energy density as a function of the chemical potential.}
\label{energia}
\end{figure}

\begin{figure}[tbh]
\vspace{0.5cm} \centerline{\psfig{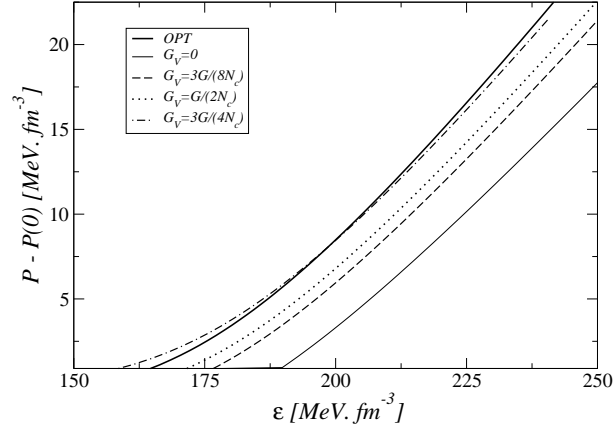}}
\caption{Equation of state for quark matter (pressure as a function of
  energy density).}
\label{EoS}
\end{figure}

The pressure subtracted by the $\mu=0$ value, $P(\mu)-P(0)$,  as a
function of the chemical potential is shown in
{}Fig. \ref{pressao}. We also see from the results  that the OPT
prediction is very close to the MFA with $G_V=G/(2N_c)$ near
criticality, but the OPT curve is less steep than the MFA one and
tends to  deviate  towards  higher values of $G_V$ as $\mu$ increases.
Thus, one may expect that the baryon density predicted by the OPT to
be smaller at high $\mu$. This expectation is confirmed by the results
shown in {}Fig. \ref{densidade}, where we show the baryon density as a
function of the chemical potential.  The same behavior is also
reflected in {}Fig. \ref{energia}, where is shown the energy density
as a function of $\mu$. Again, one sees that the OPT and the MFA with
$G_V= G/(2N_c)$ are in best agreement around criticality, as shown by
the results in Table \ref{tab}, while at higher values of $\mu$ the
OPT result interpolates between this  and higher effective $G_V$
values. This behavior of the OPT results at larger chemical potential
can be predicted if we compare the OPT expression for the free energy
density, Eq. (\ref{landauenergy}), with the one in the MFA
approximation, Eq. (\ref{landauenergyMFA}). Besides of  the nontrivial
dependence of the expression, in the OPT case, with the variational
parameter $\eta$, there is also the additional contribution with the
$I_2$ term in Eq. (\ref{landauenergy}). All these contributions  from
the OPT tend to favor a comparatively larger value of $G_V$ in the MFA
case,  for higher values of the chemical potential, when trying to
match the OPT results with the ones within the MFA with $G_V \neq 0$.
In the next section we will trace this behaviour by deriving a more
analytical connection between the OPT and such an effective mean field
vector-vector coupling.  {}Finally, this behavior of the OPT compared
with the MFA case with a non vanishing value of $G_V$, also reflects
on the EoS. The EoS results for the different cases are shown in
{}Fig. \ref{EoS} and serve to illustrate how the OPT corrections
produces a stiffer equation of state for this  version of the
model \cite{njlsu2}.

\section{A simple analytical approximation}
\label{sec4}

We shall try to get a deeper insight about our previous numerical
results by examining a simple well-motivated approximation. As we will
see, it will essentially explain the bulk of the OPT results versus
the MFA ones with extra vector coupling,  explaining in particular the
rather accidental best OPT matching by   the HF $G_V=G/(2N_c)$ value
near criticality.   {}First, as it is well-known, the large-$N_c$ (or
MFA) is  also equivalent to the  traditional Hartree approximation,
which only incorporates effects from direct (tadpole) contributions,
while exchange terms are also considered within HF.  Indeed, the value
$G_V=G/(2N_c)$ is just the vector coupling one obtains by a {}Fierz
transformation of the original interaction, Eq.~(\ref
{Lnjl}). Therefore,  Eq.~(\ref {landauenergyMFA}) with this value of
$G_V$ just corresponds to the HF approximation.  Similarly, the
two-loop induced first order OPT corrections to the free energy
incorporate the structure given by $1/N_c$ exchange-graph of the
{}Fierz type. Indeed for $\mu=0$ (and $T=0$) the OPT results are
generically consistent\cite{njlsu2} with HF ones (up to higher
$1/N^2_c$ corrections originating from the optimization of the mass
$\eta$).   Note however that in the Abelian case, $M_{\rm HF}$ is
identical to  $M_{\rm MFA}$, due to the exact cancellation of the
$1/N_c$ scalar and  pseudo-scalar contributions.   However, as far as
the $\mu$-dependence is concerned, OPT corrections also induce extra
terms, as is already clear from Eq.~(\ref{simplePMS0}). Comparing the
set  of equations above one sees that nonperturbative  information
concerning a vector type of interaction is being taken into account by
both approximations  in a different way. This becomes clear if one
recalls that the $I_3$ integral roughly  represents the quark number
density, $\langle \psi^+ \psi \rangle$, while the integral $I_2$
roughly represents the scalar condensate, $\langle {\bar \psi} \psi
\rangle$. The last term of  Eq.~(\ref {simplePMS0})  explicitly
displays that $I_3$ is also considered in a nonperturbative fashion.

In order to understand further beyond purely numerical results  the
origin of these differences, let us have a deeper look at the
expressions for the MFA and OPT mass gap expressions,  respectively
given by Eq.~(\ref{Mmc}) with (\ref{mushift}) for the former (where
$I_2$ and $I_3$ are  to be evaluated at  $\tilde{\mu}$), and
Eqs.~(\ref{simpleGAP}) and (\ref{simplePMS0}) for the latter.
{}For $\mu >\mu_c$ a simple but crucial observation is that  the
behavior of both expressions is essentially driven by the explicit
chiral symmetry breaking mass $m_c$. Indeed, for $m_c=0$, one would
have $M_{MFA}(G_V)(\mu >\tilde \mu_c) \equiv 0$ and  $M_{OPT}(\mu
>\mu_c) \equiv 0$ consistently. Now, since $m_c$ is small as compared
to all other scales, we may expect that to a good approximation the
mass gap could   be well approximated by a first-order expansion in
$M$, next to be   solved by a self-consistent equation (becoming in
that case a trivial first-order equation):

\begin{equation}
M\simeq m_c +\beta(\mu,\Lambda,G,\cdots) M + {\cal O}(M^3)\;,
\label{Mexp}
\end{equation}
where $\beta(\mu,\Lambda,G,\cdots)$ is the coefficient obtained from
this first-order expansion.  We also note that the next-order term in
the expansion in Eq.~(\ref{Mexp}) is of order $M^3$.

Explicitly, we obtain after some simple algebra:

\begin{equation}
M_{MFA}(G_V)\simeq m_c + \frac{N_c\,G}{\pi^2} \left(\Lambda^2-\tilde
\mu^2\right) M_{MFA}+ {\cal O}(M^3)\;,
\label{MFAsimp}
\end{equation}
where from Eq.~(\ref{mushift}),

\begin{equation}
\tilde \mu = \mu -2 \frac{G_V N_c}{3\pi^2}\tilde \mu^3 +{\cal
    O}({\tilde \mu} M^2) \;,
\label{muc}
\end{equation}
and 

\begin{equation}
M_{OPT}\simeq m_c + \frac{N_c\,G}{\pi^2}
\left[(\Lambda^2-\mu^2)+\frac{2}{3N_c}\frac{\mu^4}{(\Lambda^2-\mu^2)}
  \right] M_{OPT}+ {\cal O}(M^3)\;,
\label{OPTsimp}
\end{equation}
respectively. The gap equation solutions are thus trivially given as
 
\begin{equation}
M_{OPT}= m_c (1-\beta_{OPT}(\mu,\Lambda,G,\cdots))^{-1}\;,
\label{solsimp}
\end{equation}
and similarly for $M_{MFA}(G_V)$. Note that those approximations both
turn out to be accurate at the percent level, as compared to the exact
numerical solutions, for a large range of $\mu_c \lesssim \mu \ll
\Lambda$ values.  This is largely due to the very suppression by
higher $1/(\Lambda^2-\mu^2)$ powers of higher order coefficients of
these expansions in powers of $M$. 

Now defining generically $G_V = \alpha\, G/N_c$, $\alpha$ will depend
essentially on $\mu$ and this is what we would like to determine next.
Before that, we observe that, at least for rather moderate values of
$G_V \sim {\cal O}(G)$, Eq. (\ref{mushift}) when solved exactly
actually gives a moderate shift of $\tilde \mu$ with respect to $\mu$,
giving $\tilde \mu$ slightly smaller than $\mu$ by only a few percent.
So $\mu_c$ is well approximated by restricting Eq. (\ref{muc}) at
first iteration, i.e., first $G_V$ order.  This allows to derive a
simple relation to determine the relevant coefficient $\alpha$ above
for the $G_V/G$ ratio, as a function of $\mu$ and the other
parameters. {}From the matching $M_{MFA}=M_{OPT}$, then from Eqs.~(\ref{MFAsimp})
and (\ref{OPTsimp}) and using Eq.~(\ref{muc}) at
first iteration, but solving exactly the (quadratic) equation
for $\alpha$, after a simple algebra one finds

\begin{eqnarray}
\alpha &=& 
\frac{3 \pi^2}{2 G \mu^2} \left[ 1-\sqrt{1-\frac{2\mu^2}{3 N_c (\Lambda^2- \mu^2)}}
\right]\nonumber \\
&=&
\frac{\pi^2}{N_c \, G \Lambda^2} \left[1-\frac{\mu^2}{\Lambda^2}
+\sqrt{\left(1- \frac{\mu^2}{\Lambda^2} 
\right)^2-\frac{2}{3 N_c} \frac{\mu^2}{\Lambda^2} \left(1- \frac{\mu^2}{\Lambda^2} 
\right) }\right]^{-1},
\label{alpha}
\end{eqnarray}
which gives an explicit relation for the appropriate $G_V$ values
needed to match the OPT results for arbitrary $\mu$ (only  valid of
course for $\mu \ge \mu_c$,   and also for $\mu \ll \Lambda$ in
principle, since reaching too close to the natural cutoff does not
make much sense physically within the NJL effective model).

In deriving Eq. (\ref{alpha}) it is assumed that both the OPT and MFA
masses  involve the very same current mass $m_c$, which is thus valid
if $m_c$ is a common input to both  approximations. However, as
discussed above in Section \ref{sec3}, the OPT $m_c$ value as
determined consistently from fitting $m_\pi$  is slightly smaller than
the corresponding MFA value. Although a rather modest change in $m_c$
it affects the results, due to the high sensitivity to $m_c$ for $\mu
> \mu_c$, making the OPT mass gap lower by the same amount according
to Eq. (\ref{solsimp}) above, as compared with the MFA mass gap for a
given $G_V$ value.  This partly delays the increase in the effective
$G_V$ for increasing $\mu$ from Eq.~(\ref{alpha}), which can be easily
taken into account and  combined with the latter to
extract a precise relation for $G_V/G$ appropriately modifying
Eq.~(\ref{alpha}).   Since $m_c^{OPT}/m_c^{MFA} \sim 0.93$, this gives
a moderate modification to the relation (\ref{alpha}) in practice,
making the corresponding effective $G_V$ slightly smaller  by about
10\% for a given $\mu$.  This fits remarkably well our numerical
results for the mass gap obtained with the exact expressions, with
only a few percent errors.  {}For $\mu\sim \mu_c\simeq 360 \, {\rm
  MeV}$, i.e. very near criticality, one finds $G_V N_c/G \simeq
0.53$, which explains a posteriori why the results remain there very
close to the HF value $G_V N_c/G \equiv 1/2$. However, it is also
clear from the previous derivation that this is to some extent a
numerical accident,  resulting from a rather fortuitous combination of
the Abelian model factors,  $G\Lambda^2\simeq 4.44$ and
$\mu_c/\Lambda$ values such that the term in parenthesis   in
Eq.~(\ref{alpha}) is $\sim 0.78$, plus the small correction from
$m_c^{OPT}/m_c^{MFA}$ as above explained. Indeed for higher $\mu$
$G_V/G$ grows first moderately and then more rapidly when $\mu$
increases towards $\Lambda$.  {}For example, we obtain $G_V(\mu\sim
400\, {\rm MeV})\sim 0.6 G/N_c$ and $G_V(\mu\sim 500\, {\rm MeV})\sim
1.13 G/N_c$.  This also illustrates the differences between OPT and HF
or any other  fixed $G_V$ value, since OPT mimics an effective
$G_V(\mu)$ value. {}Finally, the behaviour of the other physical
quantities in Figs.~2-5 may be understood semi-analytically along the
same line of reasoning, though we skip the details here. The pressure
in Fig. 2 follows  roughly the same trend as the mass gap. As for the
other quantities in Figs. 3-5, the more pronouced increase of the
effective OPT-matching $G_V$ value at increasing $\mu$ values   can be
understood essentially by the enhanced $\mu$-dependence from  the
definition of the density. 

\section{Conclusions}
\label{sec5}

In order to investigate the physical origin of the main effects
produced by the OPT nonperturbative approximation, we have evaluated
the  free energy density for the Abelian NJL model.  This version of
the model is certainly less realistic than the non-Abelian ones,  but
being much simpler, it suits our purpose by providing an easier to
analyze free energy.  As usual, in the large-$N_{c}$ limit, the OPT
reproduces the MFA result exactly,  while at the next-to-leading
$1/N_c$ order as induced by OPT, the contributions from the scalar and
pseudo-scalar channels cancel each other whenever $\mu =0$ (for this
version of the model \cite{njlsu2}).  Nevertheless, the finite $N_{c}$
contributions may still be important for situations such as $T=0,\mu
\neq 0$ (dense cold quark matter) and $T\neq 0,\mu \neq 0$ (hot and
dense quark matter). 

Considering the MFA predictions for the same version of the model one
observes that the  OPT prediction for the critical chemical potential
value ($\mu_c$), at $T=0$, is higher. However, as it is well known,
the MFA applied to the NJL in the presence  of an extra repulsive
vector-vector interaction (proportional to the coupling $G_{V}$) also
predicts higher $\mu_c$ values, as compared to  the $G_V=0$ case.
Then, comparing the mathematical structure of the free energy density
provided by the OPT, with $G_V=0$, and  the MFA, with $G_V \ne 0$, we
show  that the former approach radiatively generates an effective
vector-like type of contribution which is $1/N_c$ suppressed and,
hence, does not appear in a large-$N_c$ type of  calculation. This
comparison hinted to the value $G_V = G/(2 N_c)$ as the one for which
both approximations could furnish similar results in a reasonable
range of $\mu$ values near criticality.  This has been successfully
verified in both numerical and analytical manner,  showing that the
OPT predictions, obtained with one less parameter ($G_{V}$), point out
in the same direction as the ones produced by the MFA applied to the
NJL with $G_{V}\neq 0$ and, thus, with a larger parameter space.  In
particular, the agreement for  the value of $G_{V} = G/(2N_c)$  is
best near criticality, but we have shown by a simple analytical
approximation how OPT  incorporates corrections for higher $\mu$
matching higher $G_V$ values.  Indeed it should be remarked that the
very proximity, near criticality, of the OPT results with the  simpler
HF approximation, is not much a generic feature but a rather
fortuitous accident of  the very simple Abelian model here
considered. In the $SU(2)$ version of the NJL model, the OPT
exhibits \cite{njlsu2} more departure from the HF approximation, and in
other models also it generically captures corrections beyond HF (for
instance in Ref.~\cite{gn2rgopt} results obtained from OPT at first
order for the Gross-Neveu mass gap are very close to the full
next-to-leading $1/N$ corrections).  Indeed an interesting follow up
beyond the  scope of the present work could be to examine similarly
the correspondence between a generalized OPT-NJL SU(N) and extra
four-fermion interactions in MFA.     

The present application suggests that the OPT is able to bring new
physics features, without introducing new parameters. This conclusion
certainly deserves further investigation and we are carrying out a
similar analysis for the case of general four-fermion theories in an
attempt to estimate the consequences beyond  the MFA for the QCD phase
diagram with a special concern about the location of the  critical
point. These results will be presented elsewhere.

\acknowledgments

MBP would like to thank Kenji Fukushima for discussions.  This work
has been partially supported (MBP, ROR and ES) by Conselho Nacional de
Desenvolvimento Cient\'{\i}fico e Tecnol\'ogico (CNPq) and by
Coordena\c c\~{a}o de Aperfei\c{c}oamento de Pessoal de N\'{\i}vel
Superior (CAPES).


\begin{thebibliography}{99}

\bibitem{njl} Y. Nambu and G. Jona-Lasinio, Phys. Rev. \textbf{122}
  (1961) 345;  \textbf{124}  (1961) 246.

\bibitem{klevansky} S. P. Klevansky, Rev. Mod. Phys. \textbf{64}
  (1992) 649.

\bibitem{njlsu2} J.-L. Kneur, M. B. Pinto and R. O. Ramos, Phys.
  Rev. C \textbf{81}  (2010) 065205.

\bibitem{strong} L. Ferroni, V. Koch and M.B. Pinto,  Phys.  Rev. C
  \textbf{82}  (2010) 055205.

\bibitem{prdgn2d} J.-L. Kneur, M. B. Pinto and R. O. Ramos,
  Phys. Rev. D \textbf{74}  (2006) 125020.

\bibitem{prdgn3d} J.-L. Kneur, M. B. Pinto, R. O. Ramos and E. Staudt,
  Phys.  Rev. D \textbf{76}  (2007) 045020; Phys. Lett. \textbf{B567}
  (2007) 136.

\bibitem{linear} A. Okopinska, Phys. Rev. D {\bf 35}  (1987) 1835;
  M. Moshe and A. Duncan, Phys. Lett. {\bf B215}  (1988) 352.

\bibitem{pms} P. M. Stevenson, Phys. Rev. D {\bf 23}  (1981) 2961;
  Nucl. Phys.  B {\bf 203}  (1982) 472.

\bibitem{buballa} M. Buballa, Phys. Rep. \textbf{407}  (2005) 205.

\bibitem{buballa_stab} M. Buballa, Nucl. Phys. A {\bf 611}   (1996)
  393.

\bibitem{koch} V. Koch, T. S. Biro, J. Kunz, and U. Mosel,
  Phys. Lett. B {\bf 185}  (1987) 1.

\bibitem{Fukushima1}K. Fukushima, Phys. Rev. D {\bf 77}   (2008)
  114028.

\bibitem{Fukushima2} K.~Fukushima, Phys.\ Rev.\ D {\bf 78} (2008)
  114019.

\bibitem{ederson} E. Staudt, Ph.D Thesis, UFSC, 2009
  (www.pgfsc.ufsc.br).

\bibitem{gn2rgopt} J.-L.~Kneur and A.~Neveu, Phys. Rev. D {\bf 81}
(2010) 125012.

\end{thebibliography}
\end{document}